\documentclass[aps,prl,reprint,groupedaddress,nobibnotes,showpacs]{revtex4-1}

\usepackage{graphicx}

\begin{document}

\title{Measuring node spreading power by expected cluster degree}

\author{Glenn Lawyer}
\email[]{lawyer@mpi-inf.mpg.de}
\homepage[]{mpi-inf.mpg.de/lawyer}
\altaffiliation[see also: ]{www.scipirate.com}
\affiliation{Max Planck Institute for Informatics}

\date{\today}

\begin{abstract}
Traditional metrics of node influence such as degree or betweenness identify
highly influential nodes, but are rarely usefully accurate in quantifying the
spreading power of nodes which are not. Such nodes are the vast majority
of the network, and the most likely entry points for novel influences, be 
they pandemic disease or new ideas. Several recent works have suggested
 metrics based on path counting. The current work proposes instead
 using the expected number of infected-susceptible edges, and shows
that this measure predicts spreading power in discrete time, continuous time,
and competitive spreading processes simulated on large random networks
and on real world networks. Applied to the Ugandan road network, it predicts 
that Ebola is unlikely to pose a pandemic threat.
\end{abstract}

\maketitle

Networks have become the premier approach to describing spreading processes 
such as epidemics because they express the heterogeneity of interactions 
\cite{Danon2011}. Early metrics of node influence focused on 
identifying highly influential nodes from the macroscopic structure of the 
network such as  degree \cite{Albert2002}, k-shell \cite{Kitsak2010},
 or centrality \cite{Freeman1979,Friedkin1991}.
These measures, however, only rank the nodes without quantifying the outcome 
\cite{Bauer2012}, and do not account for the dynamics of the spreading process 
\cite{Klemm2012}. 

Highly influential nodes are unlikely to be disease entry points. They are, by 
definition, rare.  Nor are they biological
 targets. More than half of all new or emerging infectious disease agents
 in humans are zoonotic in origin \cite{Taylor2001,Reperant2010} and thus
 closer to the periphery of society. Highly contagious diseases such as 
pandemic influenza circulate at low levels for months or years before epidemic 
breakout \cite{Xu2012}. Worryingly, structural measures of node
centrality may considerably underestimate the spreading power of 
non-hub nodes \cite{Sikic2011}.

Only recently have measures been proposed which take into account the spreading 
process. To date, and to the best of our knowledge, these are limited to path 
counting approaches.
Path counting was first proposed as the \emph{Accessibility} metric, the 
exponential of the entropy of the number of paths  
\cite{Travencolo2008,Viana2012}.
Path counting is also the basis of the \emph{impact} \cite{Bauer2012} 
and the \emph{dynamic influence} \cite{Klemm2012},  
both of which include transition/transmission probabilities when calculating 
path length.
A recent comparison of a number of measures, including accessibility, 
betweenness centrality, clustering, degree, and k-shell, found that the  
accessibility and the (weighted) degree were most predictive of an individual 
node's spreading potential across a range of different network models 
\cite{daSilva2012}.
Spreading processes, however, are not constrained to follow paths; they form 
clusters.

We here show that measuring node spreading power by taking the 
expectation of the degree of a disease cluster seeded from a single node
accurately quantifies the spreading power of that node.
Spreading power is assessed in three contexts. The simplest epidemic model is
 a susceptible-infected (SI) epidemic. Since infected nodes do not 
recover, such a process will in time infect every node connected to a seed 
node. 
When transmission is modelled in continuous time,  node spreading power can
 be measured in terms of the expected time until half the network is covered.
A model allowing for recovery,  the susceptible-infected-susceptible (SIS), 
raises the possibility that the outbreak dies out if nodes do not transmit
 before recovery. Node spreading power can here be measured by estimating a 
node's probability of seeding an epidemic. 
Node spreading power can thirdly be assessed via a competitive spreading 
process in which two mutually hostile infections invade a network.
This problem has been attracting considerable attention in the alogirthmics 
and social networks communities, where it is now known that determining the
 optimal starting point(s) is NP-hard \cite{Kostka2008}. 
In scale-free networks, where growth in prevalence is near instantaneous
 \cite{Lloyd2001,Barthelemy2005}, with asymptotic time to full coverage 
$\log (\log (n))$ \cite{fountoulakis2012}, victory goes to the team which is
nearer to instantaneous. The relative spreading power of the seed nodes 
determines the outcome.
We conclude by assessing the spreading power of nodes formed by
junctions in the Uganda road network, in the context of the recent Ebola 
outbreaks.

\begin{figure}
 \includegraphics[width=0.5\textwidth]{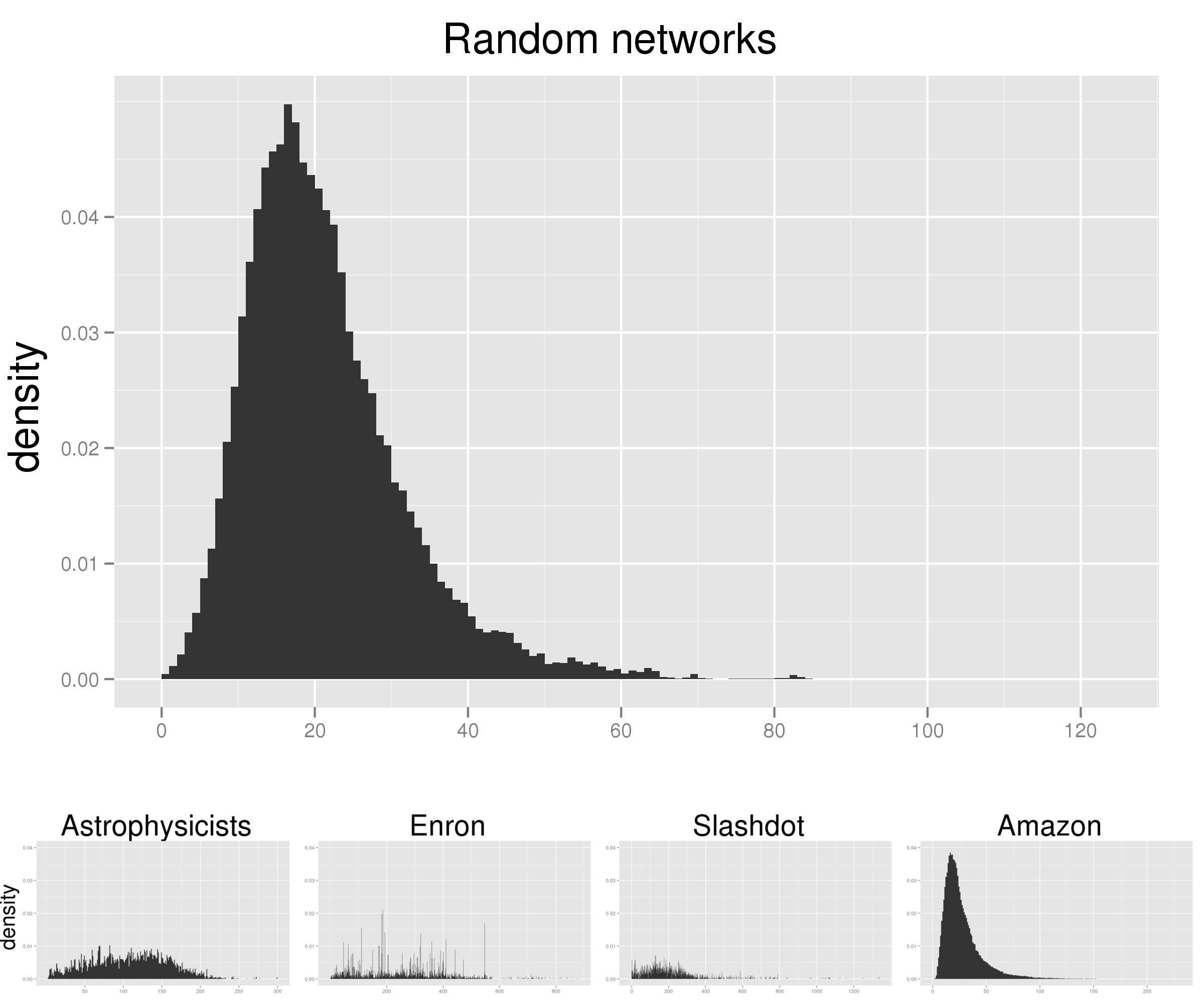}%
 \caption{\label{fig:erhist} (Top) Histogram of $ER_3$ values from the 75\% 
of nodes which are three or more hops distant from a hub, taken over 10 
random networks. (Bottom) Histograms of $ER_3$ values from the four random 
networks.}
 \end{figure}

Define the degree of a cluster of nodes as the number of edges connecting nodes
 within to nodes outside the cluster. Then the \emph{Expected Reach} of node
 $i$, $ER_X(i)$, is the expectation of the degree of the infected cluster
 formed after $X$ infections seeded from  $i$.
As the rate of disease spread is inversely proportional to the number of edges 
from infected to susceptible individuals, 
it is often preferable to work with the inverse of the $ER_X$,  the 
\emph{Expected Wait} $EW_X$.
In a  continuous time processes seeded from node $i$, the expected wait until 
infection $X+1$ given that $X$ infections have occurred is
 $EW_X(i)=\beta/ER_X(i)$, where 
$\beta$ is the transmission probability along a single edge.
The distribution of $ER_3$ values shown in the histogram in 
Figure \ref{fig:erhist}.

The $ER_X(i)$ can be found by enumerating all possible clusters of infected
 nodes which could occur after $X$ infection events originating from $i$ and
taking the mean cluster degree of this set.
For $X=0$, the cluster is node $i$, and  $ER_0$ is the degree of $i$.
For $X=1$, the clusters are the set of all pairs $\{i,j\}$ where $j$ is a 
neighbour of $i$, the degree of each cluster is $deg(i,j)=deg(i) + deg(j) -2$, 
and $ER_1$ is the mean of  $deg(i,j)$ taken over all $j$ which are neighbours 
of $i$.

Results are based on extensive simulations conducted both on random scale-free
 networks with $2^{13}$ nodes and real-world networks with $2^{14}$ to $2^{21}$ 
nodes. The random networks are generated to have a Pareto (1, 2.3) degree
 distribution under the Chung Lu protocol \cite{Chung2002}. 
Real world networks include:
the collaboration network from ArXiv Astrophysics \cite{Leskovec2007},
Enron emails  \cite{Leskovec2009},
the Slashdot Zoo signed social network from Feb 21 2009 \cite{Leskovec2009}, and
Amazon co-purchases  \cite{Leskovec2007a};
see Table \ref{tab:realnets}.

\begin{table}[h]
\caption{\label{tab:realnets}  Characteristics of the random and real networks
including the number of nodes, largest eigenvalue $\alpha$, and graph density.}
\begin{ruledtabular}
\begin{tabular}{lllrl}
                                 &   nodes & $\alpha$ & density\\
Random                           &   8,192 &  12.2 &  4.32 e-04\\  
Astrophysics \cite{Leskovec2007} &  18,772 &  94.4 &  0.22 e-04\\
Enron  \cite{Leskovec2009}       &  36,692 & 118.4 &  2.73 e-04\\
Slashdot \cite{Leskovec2009}     &  82,168 & 124.7 &  1.61 e-04\\
Amazon   \cite{Leskovec2007a}    & 262,111 &   5.3 &  0.26 e-04\\
\end{tabular}
\end{ruledtabular}
\end{table}

The $ER_3$ of all peripheral nodes is measured, with peripheral nodes defined 
as those three or more hops distant from the closest hub, 
and hubs defined as nodes with degree greater than 60\% of the maximum degree 
node of the network.
Approximately 75\% of the nodes in the random networks meet this criteria.
The Slashdot and Amazon networks have over $2^{16}$ nodes; in these cases only 
nodes greater than three hops from a hub are considered.
In the random networks, $ER_3$ is quantinized to $EW_3$ by 
truncating the inverse to the next lower hundredth, i.e.
 $ER_3 \in (20,25] \rightarrow EW_3 = 0.04$.
This gives approximately equal number of nodes for each value of $EW_3$.
In the real-world networks, the quantization is independently scaled 
give approximately uniform representation of the lower values of the resulting 
$EW_3$.

Expected reach is predictive of the mean time for an
 infection originating at a peripheral node to cover half the network
 in a disease without recovery simulated in continuous time (SI model).
Expected time to half coverage (tthc) is measured by simulating $2^7$ epidemics 
 for each seed node, measuring the time until half the nodes 
are infected, and fitting the measurements to an exponential distribution.
Seed nodes are five randomly selected nodes at each observed value of $EW_3$ 
on $2^5$ random networks.
For higher (and thus rarer) $EW_3$ values, it is not always possible to select 
five nodes; in such cases all observed $EW_3$ values are used. 
The accessibility metric \cite{Travencolo2008} is also measured for each node.
Logistic regression over the resulting 4563
observations shows that both $EW_3$ and accessibility are highly predictive of 
tthc, with longer expected wait associated with longer tthc.
The $EW_3$, however,  explains more of the deviance (71\% vs 48\%) and has a 
lower AIC (-12376 vs -9831).
A similar procedure was performed on the real-world networks, again achieving
 significance in all cases as detailed in  Table \ref{tab:realfits}.

\begin{figure}
 \includegraphics[width=0.5\textwidth]{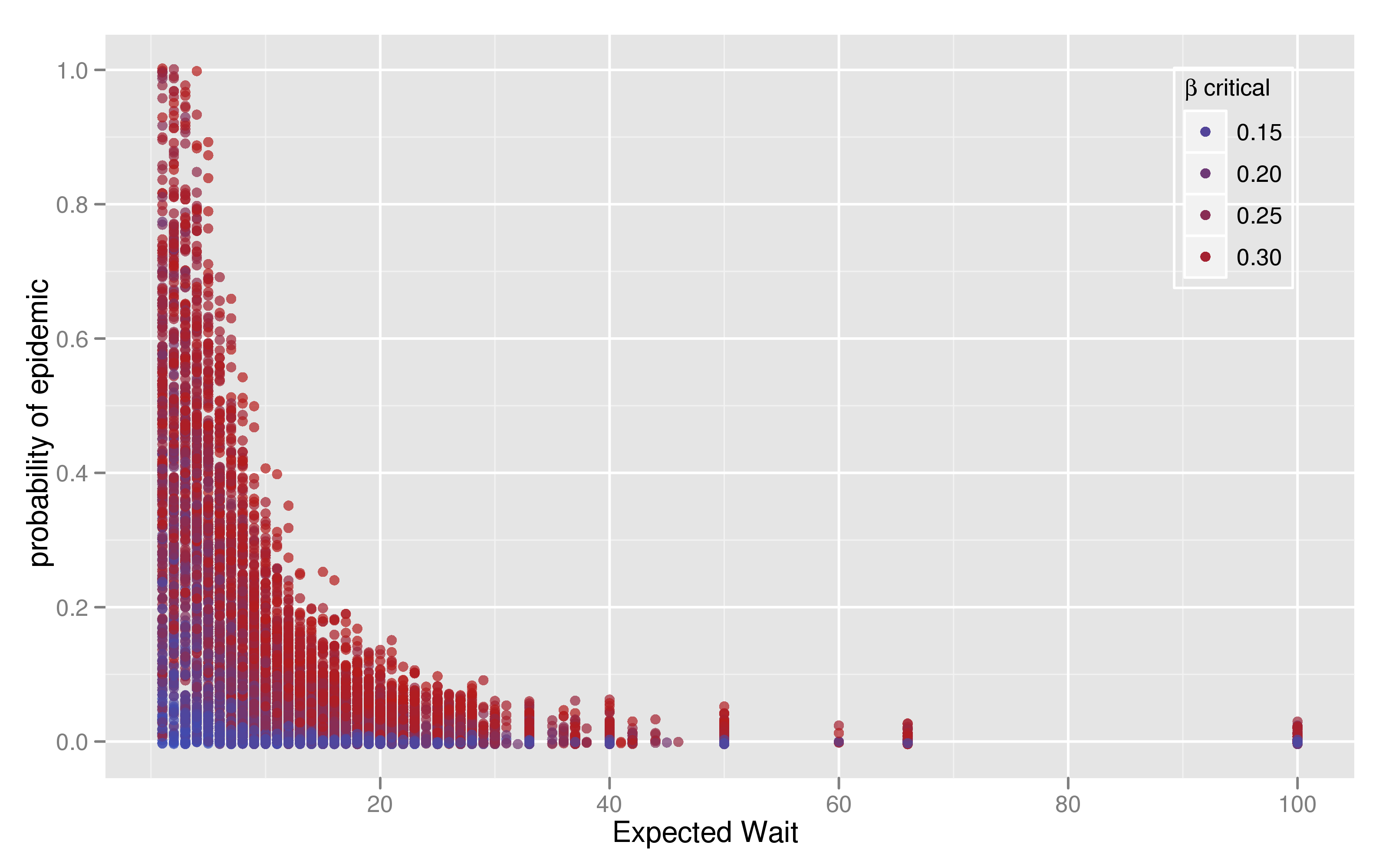}%
 \caption{\label{fig:pep} (Color online) In an SIS model, disease which cannot
 rapidly establish itself in the population dies out. A node's  $EW_3$ is 
associated with its ability to seeding an epidemic. Disease transmissibility,
 $\beta$, is set to five times the critical value separating the 
endemic/extinction regimes.}
 \end{figure}

The expected reach is indicative of a node's ability to seed a sustained 
epidemic in disease with unit time recovery simulated in discrete time  
(SIS model). 
Epidemic potential is measured by simulating 100 epidemics and counting how
 many persist for 50 iterations \footnote{In 20620 runs, infections which go 
extinct usually (93\%) do so in less than five iterations, and universally in 
less than 20.}.
 Klemm et al.\ propose that for a discrete
 time process with unit recovery, the critical transmission probability
 $\beta$ value  separating the extinciton from endemic regime is the
 inverse of the largest eigenvalue $\alpha$ of the adjacency
 matrix  \cite{Klemm2012}. This suggests that $\alpha$ is also a measure of a
 network's susceptibility to infection. 
Transmision probability is here set to $\beta=5/\alpha$, placing the simulated
 epidemic well within the epidemic regime, with  $\alpha$  determined
 independently for each random network.
Five seed nodes are sampled for each observed $EW_3$ value on $2^6$ random 
networks
 A generalize additive model of the resulting  8678 observations shows that 
the combination of $ER_3$ and $\beta$ (thus controlling for the network's 
susceptibility to epidemic) explain 68\% of the deviance in the probability
 that a node can start an epidemic. 
Due to the large number of observations at low $ER_3$, the relationship is 
more clearly illustrated in terms of the expected wait (Figure \ref{fig:pep}).
The $ER_3$ also explains a significant portion of the deviance in the 
real-world networks, as shown in Table \ref{tab:realfits}. 
Setting $\beta=5/\alpha$ did not produce consistent results in the real world
 networks. No nodes had high epidemic potential at this level, with the 
exception of the Amazon network in which all did. To control for this 
variability, $\beta$ is tuned such that nodes with high $ER_3$ have epidemic
 potential approaching 100\%.  In the astrophysics collaboration network, 
this requires setting $\beta= 14/\alpha$, reducing the model fit from
 explaining 53\% (at $\beta=5/\alpha$) to 47\% of the deviance.

\begin{table}[h]
\caption{\label{tab:realfits}  Expected reach explains a significant 
percentage of the deviance in node time to half coverage (SI model) and
 probability of seeding an epidemic (SIS model) in real world networks.}
\begin{ruledtabular}
\begin{tabular}{lclcl}
             &\multicolumn{2}{c}{SI model}     &\multicolumn{2}{c}{SIS model}\\
             &dev explained & p-value &  dev explained & p-value \\
Astrophysics & 47\% & 1.10e-07 & 47\% & $<2^{-16}$ \\ 
Enron        & 38\% & 9.95e-06 & 15\% & 1.2 e-6\\ 
Slashdot     & 32\% & 0.037    & 17\% & 1.24 e-4\\ 
Amazon       & 65\% & 0.0005 & 80\%& $<2^{-16}$\\ 
\end{tabular}
\end{ruledtabular}
\end{table}

\begin{figure}
 \includegraphics[width=0.5\textwidth]{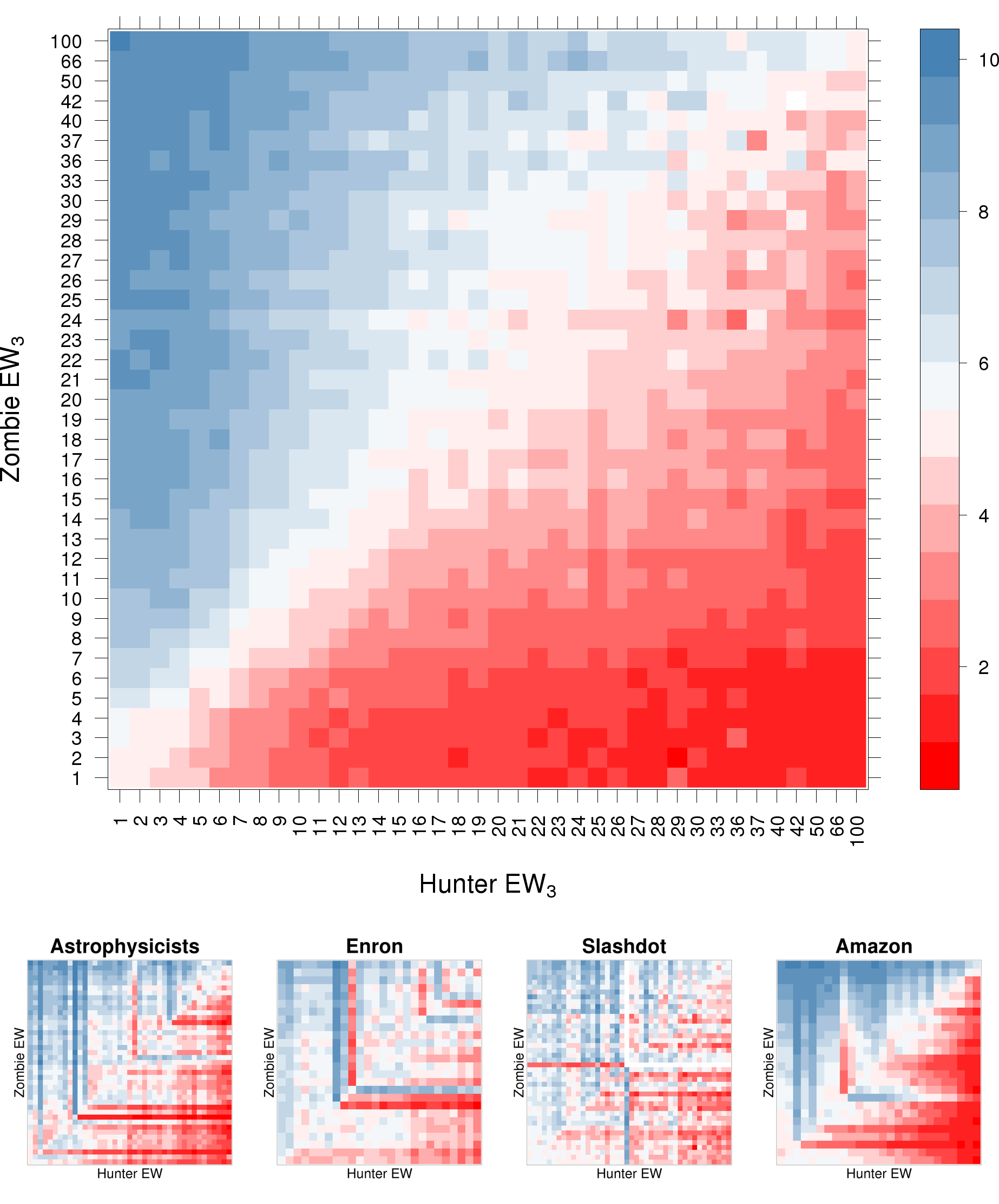}%
\caption{\label{fig:competitive}  (Color online) The winning side in the
 competitive rumor spreading process is accurately predicted by $EW_3$;
 the side with the shorter expected wait wins at least 50\% of the time,
 with greater difference giving greater advantage. The heatmap shows how 
many trials out of ten are won by the humans, with points on the grid
 indicating the $EW_3$ of the initial zombie and hunter node. Blue is good, 
red is bad. Upper plot: random networks; Lower plot: real world networks.}
\end{figure}

The expected reach predicts the outcome of a competitive spreading process
 simulated in continuous time. 
Here, we use a toy example of a zombie apocalypse met by concurrent spread of 
education in zombie hunting. Both zombies and hunters recruit from the 
susceptible population and mutually eliminate each other. 
For each network, a grid is formed for all 
possible pairings of observed $EW_3$ values. 
Ten epidemics are simulated at each point in the grid with one initial zombie 
and one initial hunter chosen randomly from nodes with $EW_3$ values 
corresponding to the grid point value.
The apocalypse is simulated in continuous time with the base rate of 
zombification and of hunter training equal. 
The outcome measure is the number of times (out of ten) the humans win, 
averaged at each point of the grid  over $2^7$ random networks.
A analogous process is applied to the real-world networks.
In the random networks, the spreading process with the lower expected 
waiting time wins at least 50\% of the time.  
The larger the difference in $EW_3$, the greater the margin of victory.
Results are similar, though less distinct, in the real world network 
(Figure \ref{fig:competitive}.)

\begin{figure}
\includegraphics[width=0.45\textwidth]{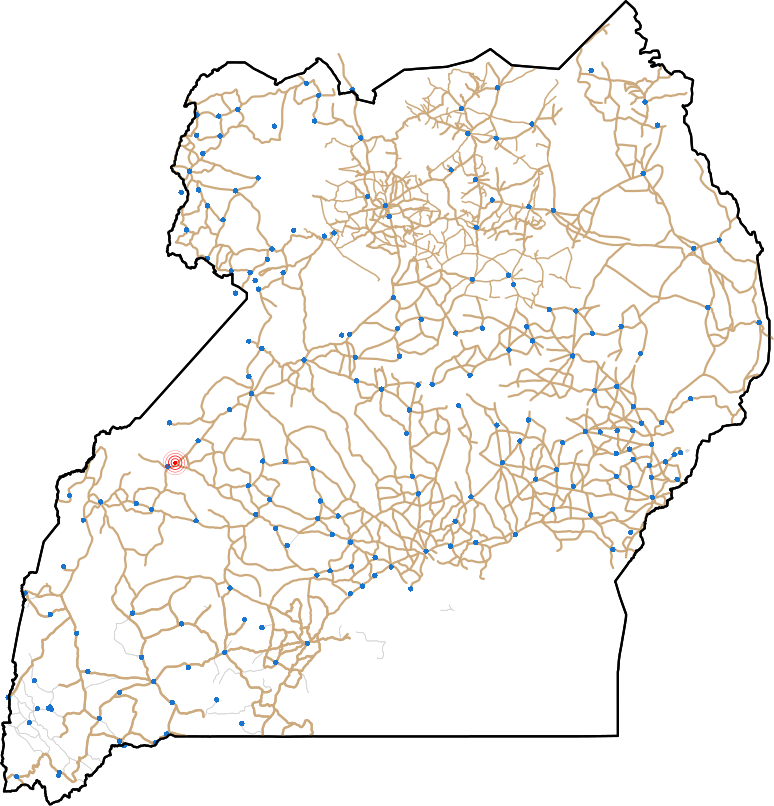}%
\caption{\label{fig:ugroads} (Color online) The Uganda road network. 
Towns are blue dots. The red bullseye marks the village of Kigadi.} 
\end{figure}

Finally, we turn our attention to the recent Ebola outbreak in Uganda.
We make the simplifying assumption that the disease transmits between 
communities along the road network, and we regard this network as an 
unweighted, undirected graph with nodes at each road junction.
Road data current as of 2009 is available from Humanitarian Response 
\cite{humres}.
The 2012 epidemic originated in the village of Kigadi %\cite{msfpress},
which sits at the juncture of three roads (Figure \ref{fig:ugroads}). 
The $ER_3$ of this juncture is 5.1, implying that it has limited spreading 
power. 
In fact, the majority of peripheral nodes have $ER_3<6$, suggesting that 
Elboa would have a hard time transitioning from a local to a national 
epidemic even with a less vigorous response from Medicine Sans Frontieres.

While $ER_3$ is sufficient for predictive purposes, the expectation is not 
unproblematic. The distribution over which the expectation is taken 
 is strongly bimodal. Most (~80\%) of the nodes analyzed here are three
 hops distant from a hub. The infection clusters which contain this hub
 have reach typically exceeding 300, while the 
clusters which do not have reach typically less than 100.

Computational expense is a concern, but generally not a problem in practice.
 The number of clusters grows factorially in the degree of nodes reachable 
from $i$ after $X$ hops. This problem is not severe, however, as the measure
 is designed for nodes where this degree is small, and $X=3$ is sufficient to 
determine the outcome, a result supported by the path counting literature 
\cite{Bauer2012,Klemm2012}.  Running time for our non-optimized C++ code is 
comprable to that reported by Bauer et al.\ for counting all paths of length
 4 under a SIR model \cite{Bauer2012}.

Arguements explaining that path counting is categorically different from 
degree or centrality based measures \cite{Travencolo2008} also apply to the 
expected cluster degree. In networks with heterogeneous degree distributions, 
node degree is generally not correlated with neighbours. Chained nodes with
 high betweenness centrality can lead to low expected reach. 
Expected cluster degree is also distict, though not categorically, from path 
counting in that it directly measures  the expected number of 
infected-susceptible edges.

\end{document}